\documentclass[12pt]{article}

\usepackage{amsmath,amsfonts,epsfig,color,latexsym}
\usepackage{pifont}
\usepackage{comment}

\newcommand{\be}{\begin{equation}}
\newcommand{\ee}{\end{equation}}
\newcommand{\bea}{\begin{eqnarray}}
\newcommand{\eea}{\end{eqnarray}}

\newcommand{\mE}{\mathcal{E}}
\newcommand{\gamp}{P}
\def\dbar{{\mathchar'26\mkern-12mu d}}

\let\newsection=\section
\renewcommand{\section}{\setcounter{equation}{0}\newsection}

\begin{document}

\begin{center}

{\LARGE\bf Band-aid for information loss from black holes}
\vskip 1in
\centerline{\Large Werner Israel and Zinkoo Yun\footnote{Israel:israel@uvic.ca\\\hspace*{6 mm}Yun:semiro@uvic.ca}}
\vskip .5in

\end{center}
\centerline{\large Department of physics and astronomy University of Victoria}
\centerline{\large Victoria, BC, Canada V8W 3P6}

\vskip 1in

\begin{abstract}

{\large 
We summarize, simplify and extend recent work showing that small deviations from exact thermality
in Hawking radiation, first uncovered by Kraus and Wilczek, have the capacity to carry off the
maximum information content of a black hole. This goes a considerable way toward resolving a
long-standing ``information loss paradox.''
}

\end{abstract}

\newpage

\section{ Introduction}\label{sbk1}
\hspace*{6 mm}
Hawking's dramatic 1974 announcement that black holes evaporate thermally has led to a much-discussed
paradox. If, as indicated by Hawking's semi-classical analysis, the radiation spectrum is 
exactly thermal (thus fully characterized by just a single parameter, temperature), 
the consequent loss of information violates a basic quantum principle, unitarity.
A number of resolutions of this puzzle have been proposed, some of them are quite 
radical \cite{gid1}.\\

A conservative option, developed from 1994 by Frank Wilczek and his collaborators  \cite{gid2},
is that back-reaction of the emissions on the geometry  and thermal properties of the
hole could result in information-carrying departures from thermality. Specifically,
emission of each discrete quantum is attended by a jump of Hawking temperature,
hence no longer definable by a single parameter. 
Correlations carried by the resulting deviations from thermality, while individually small,
could build up to a substantial information capacity over the course of the hole's lifetime.

Although earlier studies cast doubt on this possibility  \cite{gid3}, recent work by
Zhang et al \cite{gid4} has given it credence. They concluded that up to
$\exp S_{BH}$ bits of information can be carried off in the correlations,
which can include all of the information in the hole if, as seems reasonable, the Bekenstein-Hawking entropy $S_{BH}=A/4$ is a 
measure of the hole's \textit{information capacity}, in the sense that $\exp S_{BH}$ is 
the maximum number of bits that can be accommodated in a black hole formed by an
astrophysical collapse.

Kraus and Wilczek's key result \cite{gid2} was their arithmetic-mean prescription 
(AMP, cf (\ref{elcct22}) below) for the effective action (including back-reaction) of
a massive particle (modelled as a spherical shell in the s-wave approximation)
tunneling out of a spherical black hole. Their analysis was encumbered by use of the
intricate machinery of spherisymmetric Hamiltonian gravity and required a dozen pages of 
closely reasoned argument even for the simplest example of uncharged (Schwarzschild) evaporation.

The much simpler and more transparent treatment introduced here calls upon the 
general-relativistic dynamics of thin shells, together with the analytic properties of 
Schwarzschild's time co-ordinate $t$ over the extended Kruskal manifold, and occupies just a few 
lines (Sec. \ref{sbk4}). Moreover, this extends immediately to charged evaporation and 
brings new aspects of the problem into focus -- breakdown of AMP when interactions
with nongravitational forces are introduced, the extremal (zero-temperature) limit
and the possibility of black hole remnants. These aspects are further discussed
(but not fully resolved) in Sec.  \ref{sbk4}, and the need for further work emphasized.

By way of introduction to this somewhat novel treatment of the black hole tunneling 
problem, Sec. \ref{sbk3} rederives the classic Schwinger formula for charged pair
creation by an electric field, using the same (essentially geometrical) approach.
Comparison is instructive, revealing both the resemblance and at least one sharp
difference between the two cases.

Sec. \ref{sbk5} briefly reviews the key question on which the current literature makes
a confused impression -- does Hawking radiation with Kraus-Wilczek deviations from
thermality have information-holding correlations? -- and reaffirms the positive answer
given by Zhang et al \cite{gid4}. Sec. \ref{sbk6} concludes the paper with some open
questions.

\section{Tunneling}\label{sbk2}
\hspace*{6 mm}
``Tunneling'' (or ``barrier penetration'') is the term commonly used for energy-conserving 
quantum transitions that are classically forbidden. In such cases, a path-integral 
evaluation of the transition amplitude is hampered by the fact that the sum-over-paths is
not dominated by any single(real) path. The familiar remedy is analytic continuation of the
time coordinate to pure imaginary values. In the resulting Euclidean-signature spacetime
a classical tunneling route often exists and dominates the sum-over-paths. The procedure is 
analogous to the steepest-descent method for evaluating a real definite integral by diverting 
the integration contour through a saddle point in the complex plane. (cf \cite{gid8})

Equivalently, one can set out from the time-independent Schr\"{o}dinger equation for a fixed
energy. The ansatz $\psi=\exp iW$ yields the Hamilton-Jacobi equation for $W$ in WKB approximation.
This identifies $W$ as the Jacobi action or Hamilton's characteristic function:
\begin{equation}
W=\int (L+H)dt.
\label{tu1}
\end{equation}
Among orbits of given energy between two given configurations, $W$ is minimized by the classical
orbit. For configurations separated by a barrier, but connectable by a classical tunneling orbit
in imaginary time, $W$ becomes complex and gives a tunneling probability
\begin{equation}
\gamp=\psi\psi^\ast=\exp(-2\textrm{Im } W)
\label{tu2}
\end{equation}
in lowest-order approximation. (For more detatails, see e.g. \cite{gid9})

Equation (\ref{tu2}) has rather general validity and is readily applicable in situations 
where a tunneling path can be clearly defined, as in the case of a particle.
For a field, a reduction procedure is needed to describe the tunneling of a 1-particle excitation.
In an s-wave approximation on a spherical background (e.g. a spherical black hole) the excitation 
can be modelled as a spherical shell. (The reduction is spelled out in some detail by Kraus and
Wilczek \cite{gid10} for the special case of a charged scalar field propagating on the classical
background of a radial electric field.) When we treat tunneling out of a spherical black hole in
Sec.\ref{sbk4} we shall circumvent this rather elaborate reduction by inserting into (\ref{tu1}) 
the natural and simplest material action that is minimized by the classical orbit of the shell.

\section{Pair creation in a uniform electric field revisited}\label{sbk3}
\hspace*{6 mm}
The intuitive picture of tunneling out of a black hole suggested by the  Parikh-Wilczek
calculation \cite{gid2} is at first glance very different from the familiar one of
charged-pair creation by a strong electric field. There is no black hole counterpart
to the finite-width potential barrier that separates the oppositely charged pair. 
In a black hole the barrier is a causal one -- the light cone that forms the future horizon.
There are nonetheless strong resemblances between the two pictures. In this section we shall 
briefly re-derive Schwinger's results \cite{gid5} to leading order in a geometrical way 
that points up the resemblance as well as highlighting an essential difference.

We consider a uniform electric field $\mE$ directed along the positive x-axis
of Minkowski space-time. The energy of a point charge $e$ moving along the x-axis is
\begin{equation}
E=m\gamma-e\mE x, \qquad \gamma=\frac{1}{\sqrt{1-v^2}}
\label{elcct1}
\end{equation}

\begin{figure}
\includegraphics{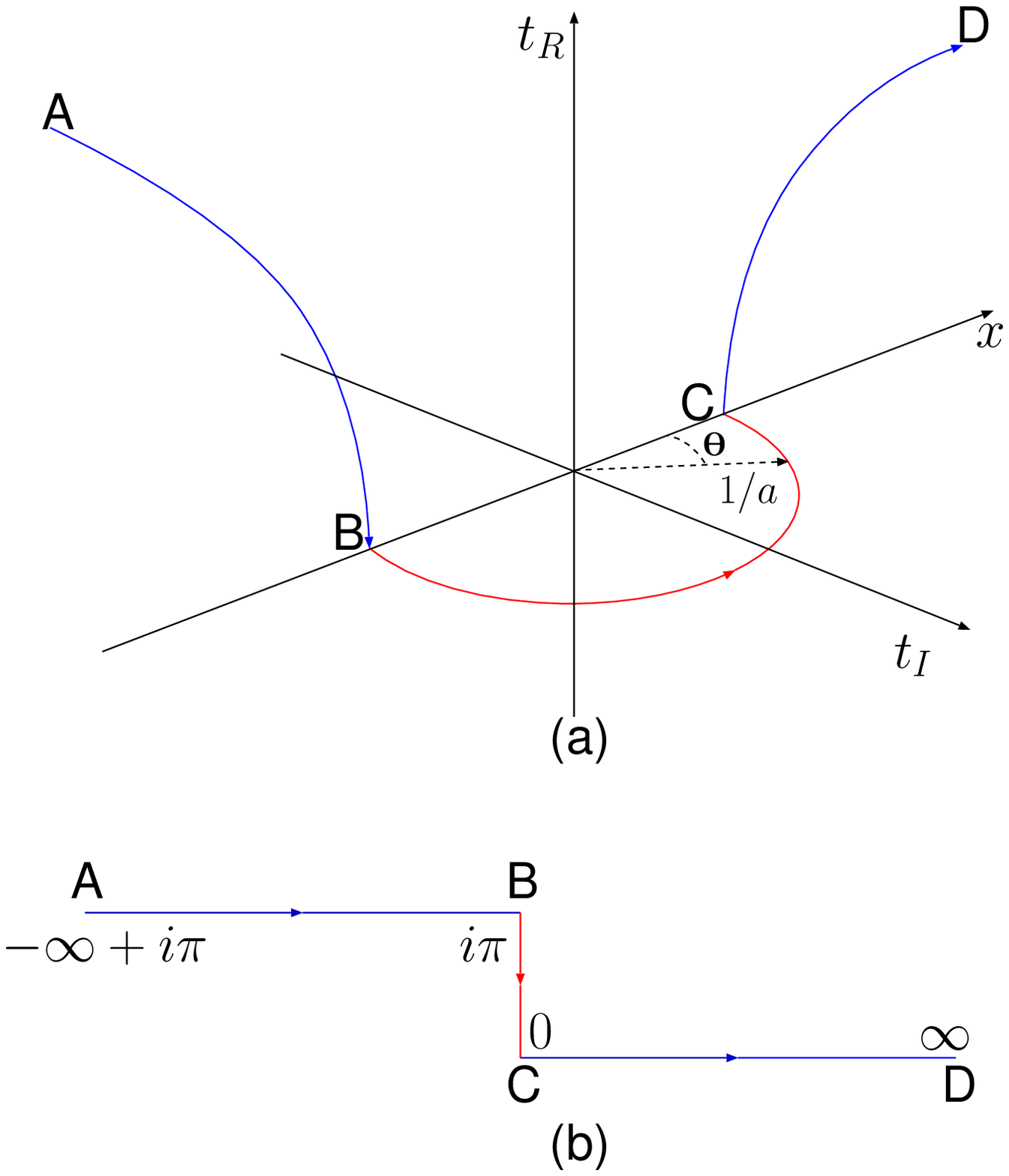}
\caption{(a) Electron pair creation as tunneling. (b) March of $a\tau$}
\label{Fig1}
\end{figure}

Intuitively, a virtual pair ($e,m$) and ($-e,m$) located at $x_+(t)$ and $x_-(t)$ can 
extract enough energy ($\geq 2m$) to emerge as a real pair if they happen to reach
a minimum separation determined by
\begin{equation}
x_+(0)=-x_-(0)=\frac{1}{a}, \qquad a=\frac{e\mE}{m}
\label{elcct2}
\end{equation}
with a convenient choice of origin. Assuming (\ref{elcct2}) as initial condition, 
the equation of motion
\begin{equation}
\frac{dp}{dt}=e\mE, \qquad p=m\gamma\frac{dx}{dt}
\end{equation}
integrates to give the hyperbolic path (parametrized by proper time $\tau$)
\begin{equation}
ax=\cosh a\tau, \qquad at=\sinh a\tau
\label{elcct3}
\end{equation}
for the newly created positive charge $e$, with $e$ replaced by $-e$ for its partner.
The two orbits are shown as CD and AB in Figure \ref{Fig1}.

It is straightforward to  recast this picture of pair creation as a one-particle
history. The particle's orbit may be traced through three stages
(AB, BC and CD in Fig \ref{Fig1} ), in each of which the orbital equation is given by 
analytic extension of (\ref{elcct2}) in the complex $\tau$-plane, $\tau=\tau_R+i\tau_I$.

In the first stage AB, given by 
\begin{equation}
-\infty \leq \tau_R \leq 0, \quad a\tau_I = \pi,
\end{equation}
so that (\ref{elcct3}) becomes 
\begin{equation}
ax=-\cosh a\tau_R, \quad at=-\sinh a\tau_R,
\end{equation}
the particle enters from the left, moving backward in Lorentz time $t$,
and is slowed by the field (so that its effective charge/mass ratio at 
this stage is really $e/(-m)$) until it comes to momentary rest at B.

The second (tunneling) stage is an excursion into imaginary time
\begin{equation}
\tau=i\tau_I, \quad a\tau_I=\theta, \quad t=it_I
\label{elcct4}
\end{equation}
and imaginary momentum $p$.
The orbit (\ref{elcct3}) is now the semicircle BC:
\begin{equation}
ax=\cos\theta,\quad at_I=\sin\theta \quad (\theta: \pi\to 0)
\label{elcct5}
\end{equation}
\\
In the third stage CD, given by  $0\leq\tau\leq\infty$, the orbit has
re-entered real time and (\ref{elcct3}) describes the conventional
picture of a particle, with charge/mass ratio $e/m$, being accelerated to
the right.

The whole sequence may be interpreted as representing a virtual charge 
$e$ which has increased its effective mass from $-m$ to $+m$ by 
extracting energy $2m$ from the field during the tunneling phase.
\\

From (\ref{elcct1}) and (\ref{elcct3}) it is easily checked directly
that the total energy  $E$ is conserved and, moreover, vanishing
for this virtual particle. Thus we do not need to distinguish
between the Jacobi and Lagrange actions in this instance:
\begin{equation}\label{elcct6}
W=I=-\int md\tau +\int eA_\alpha dx^\alpha 
=-\int(md\tau+e\varphi dt)
\end{equation}
Setting $\varphi=-\mE x$ and recalling that $e\mE=ma$,
we easily obtain from (\ref{elcct4}) and (\ref{elcct5}) for the 
tunneling phase
\begin{equation}
W^{BC}=-\frac{im}{a}\int^0_\pi \sin^2\theta d\theta=\frac{i\pi m}{2a}
\end{equation}
Hence the tunneling probability is 
\begin{equation}
\gamp=e^{-2\textrm{Im }W}=\exp\Big({-\pi m^2/e\mE}\Big)
\end{equation}
in agreement with Schwinger's classic result \cite{gid1}.

Our treatment of gravitational tunneling in the following section will closely follow these
lines, but with one key difference. In the electric tunneling phase BC, work done on the rest
mass by the field reversed its sign from $-m$ to $+m$, or (equivalently) reversed the sense
of proper time $\tau$. The effect was a sign reversal of the inertial term $-\int md\tau$ in
the action (\ref{elcct6}), which was implemented by complexifying $\tau$. In contrast, for a
particle moving under gravity the gravitational force vanishes in its rest-frame by the
equivalence principle and cannot affect its rest mass. Thus the inertial action and proper
time remain real during the tunneling phase.
 
\section{Tunneling out of a spherical black hole.}\label{sbk4}
\hspace*{6 mm}
Figure \ref{Fig2} is a Kruskal map of an eternal black hole, partitioned by the two horizon sheets(at fixed $\theta,\phi$)
into left, right, future and past sectors L,R,F and P. Our universe is taken to be the R-sector;
In terms of this idealized eternal picture, the tunneling and evaporation of a particle can come about in two
distinct ways.
\subsection{L$\to$R tunneling in an eternal black hole.}
\begin{figure}
\begin{center}
\includegraphics[height=8cm]{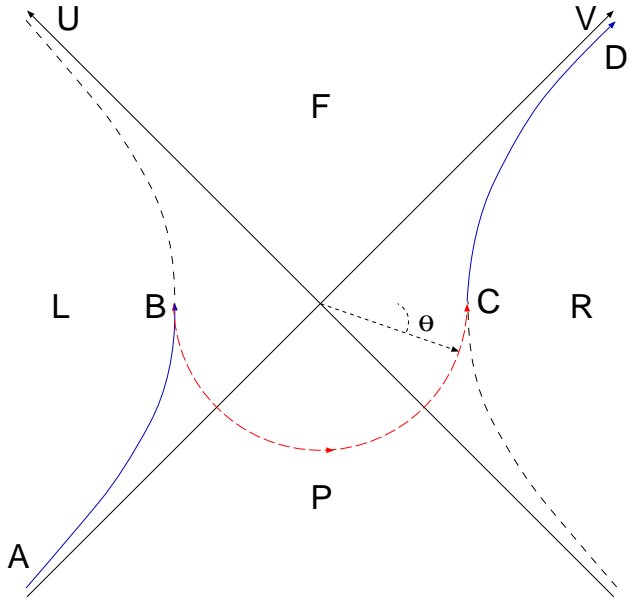}
\end{center}
\caption{L $\to$ R tunneling in an eternal black hole}
\label{Fig2}
\end{figure}

\hspace*{6 mm}
Energetically, the cheapest(hence most probable) tunneling route is to a configuration in which the
emerging particle has no kinetic energy. 
So this mechanism treats transitions between static or momentarily static configurations
on opposite sides of the horizon. A virtual particle (world line AB in Figure \ref{Fig2}) starts from rest in sector L,
enters a tunneling mode at B, then circulates in complex time around the semi-circle to C, whence it emerges in 
sector R as a real static particle. Meanwhile, the hole's mass and charge have been reduced, and the surface gravity $\kappa$
correspondingly increased, by loss of this particle - from $\kappa_L$ in sector L before the tunneling,
to $\kappa_R$ in sector R after the particle has escaped. 
This picture closely resembles the Schwinger tunneling picture of Figure \ref{Fig1}.

Eternal black holes are not formed by real astrophysical collapse. 
The initial tunneling state in sector L has no real counterpart.
So this tunneling  model is further removed from reality than the picture 
of direct tunneling from F to R that we shall consider in the next section. However, it can be formulated completely 
and precisely, and allows effects of back-reaction to be easily accommodated. The extra stretch of the
tunneling route from L into F en route to R adds no imaginary part to the action, since falling from L into F is
classically allowable.

To fix notation, the spherical geometries that we consider in this section have metrics of the general form
\[
ds^2=\frac{dr^2}{f(r)}+r^2d\Omega^2-f(r)dt^2
\]
This includes Schwarzschild and Reissner-Nordstr\"{o}m black holes imbedded in flat, de Sitter or AdS backgrounds.
At the horizon $f(r_0)=0$ and the surface gravity is $\kappa=\frac{1}{2}f'(r_0)$
(assumed non vanishing).

Advanced and retarded Eddington-Finkelstein coordinates $u,v$ are defined by
\begin{equation}\label{elcct14r}
\left\{ \begin{array}{ll}
du=dt-\frac{dr}{f(r)}\\
dv=dt+\frac{dr}{f(r)}
\end{array} \right.
\end{equation}
and the corresponding Kruskal coordinates $U,V$ by
\[
U=-e^{-\kappa u}, \qquad V=e^{\kappa v}
\]
In theses coordinates, the metric is
\[
ds^2=\frac{f(r)}{\kappa^2 UV}dUdV+r^2(U,V)d\Omega^2
\]
and is manifestly regular for all points($U,V$) where $f(r)$ is regular, including the
two horizon sheets $U=0$ and $V=0$. 

Schwarzschild time $t$ is initially defined only over the R-sector of the extended Kruskal
manifold of Figure \ref{Fig2}. It will be useful to extend it analytically to the full space 
in such a way  that $e^{-i\omega t}$ is positive-frequency with respect to the globally
regular times $U$ and $V$. 
This requires that $e^{-i\omega t}$ be regular and bounded 
in the lower halves of the complex $U$ and $V$ planes. 
A definition which achieves this is
\begin{equation}\label{elcct15d}
t=\frac{1}{2\kappa}\ln\frac{V}{U}
\end{equation}
in which we have selected that branch of $\ln z$ which is regular on the lower-half $z$-plane, and 
 real on the lower imaginary axis, so that its values,
e.g for real $x$,

\begin{equation}\label{elcct15f}
\ln x\equiv\ln|x|+\frac{i\pi}{2}\epsilon(x)
\end{equation}
are left-right symmetric. 
The imaginary constant thereby added to $t$ in (\ref{elcct15d}) does not affect 
time-differences in a given sector. Moreover, it  
ensures manifest CPT invariance of the analytically extended
transition and tunneling amplitudes that we are about to derive,  
i.e., invariance under simultaneous reversal of $t,U,V$.

Schwarzschild time $t$ is regular along the entire tunneling path ABCD(Fig \ref{Fig2}) and the background geometry is 
static (apart from small back-reaction effects). It is therefore a suitable 
candidate to be complexified and used as the parameter for L$\to$R tunneling.

In an s-wave approximation the tunneling (charged) particle is modeled as a spherical shell.
According to (\ref{A23}) in Appendix \ref{sdya}, the appropriate(Jacobi) action is
\begin{equation}
W=-\int Md\tau +\int Ed\bar{t}
\label{elcct15.5}
\end{equation}
Here, $M$ is the proper mass of the shell, $\tau$ its proper time
and the bar in $\bar{t}=\frac{1}{2}(t_++t_-)$ an arithmetical average
over the two geometries, $f_+(r)$ and $f_-(r)$ exterior and interior to the shell.
Finally, $E$ is the energy of the shell when it emerges from the tunneling phase at C.
If the shell is uncharged, and with just enough energy to reach infinity, this is simply the
Schwarzschild mass, $E=m_+-m_-$. For a shell with charge $q=e_+-e_-$ emerging  from a charged hole,
$E$ is smaller than this by $\bar{c}\equiv q\overline{(e/r_H)}$, the work done  in
lifting the shell from the horizon to infinity. Thus, generally,

\begin{equation}\label{elcct15w}
E=m-q\overline{\Phi}_H
\end{equation}
where $\Phi_H$ is the electric potential at the horizon.
From (\ref{elcct15d}),
\begin{equation}
\textrm{Im } t_B=-\frac{i\pi}{2\kappa_L}, \qquad \textrm{Im } t_C=+\frac{i\pi}{2\kappa_R}
\label{elcct15.6}
\end{equation}
valid for $t_+$, $t_-$ and $\bar{t}$.

Thus, for an emission with Schwarzschild mass-energy $E$
\begin{eqnarray}
\textrm{Im } W( E)&=&\textrm{Im}\int E d\bar{t}= E \textrm{ Im}\int d\bar{t}\label{elcct16v}
\\ 
&=&\frac{\pi}{2} E\Big(\frac{1}{\kappa_R}+\frac{1}{\kappa_L}\Big)= E\pi\overline{\left(\frac{1}{\kappa}\right)}\label{elcct16}
\end{eqnarray}
stemming from the imaginary-time path BC.
This agrees with the Parikh-Wilczek result \cite{gid2} derived in  Painlev\`{e}-Gullstrand coordinates.
Hence the tunneling probability for emission of Schwarzschild mass-energy $E$ from a charged black hole is, by (\ref{tu2}),
\begin{equation}\label{elcct17}
\gamp(E)=e^{-2\textrm{Im } W}=e^{-\beta(E) E}, \quad \beta\equiv\overline{\left(\frac{2\pi}{\kappa}\right)}
\end{equation}
in which the effective Hawking temperature $T_H=1/\beta$ is given as the harmonic mean of surface 
gravities $\kappa_L$ and $\kappa_R$ before and after emission.

The energy distribution (\ref{elcct17}) was first derived by Kraus and Wilczek \cite{gid2} by a more elaborate route.
Since $\beta$ depends on $E$, (\ref{elcct17}) is no longer a simple Boltzmann exponential. As shown by Zhang
et al \cite{gid4}(and as we shall outline in Sec. \ref{sbk5}) correlations present in 
these deviations from thermality have the capacity to carry off the maximum information content 
of the hole.

\subsection{F$\to$R tunneling:tunneling through the future horizon of an  astrophysical black hole}\label{chpms5c}

\begin{figure}
\begin{center}
\includegraphics{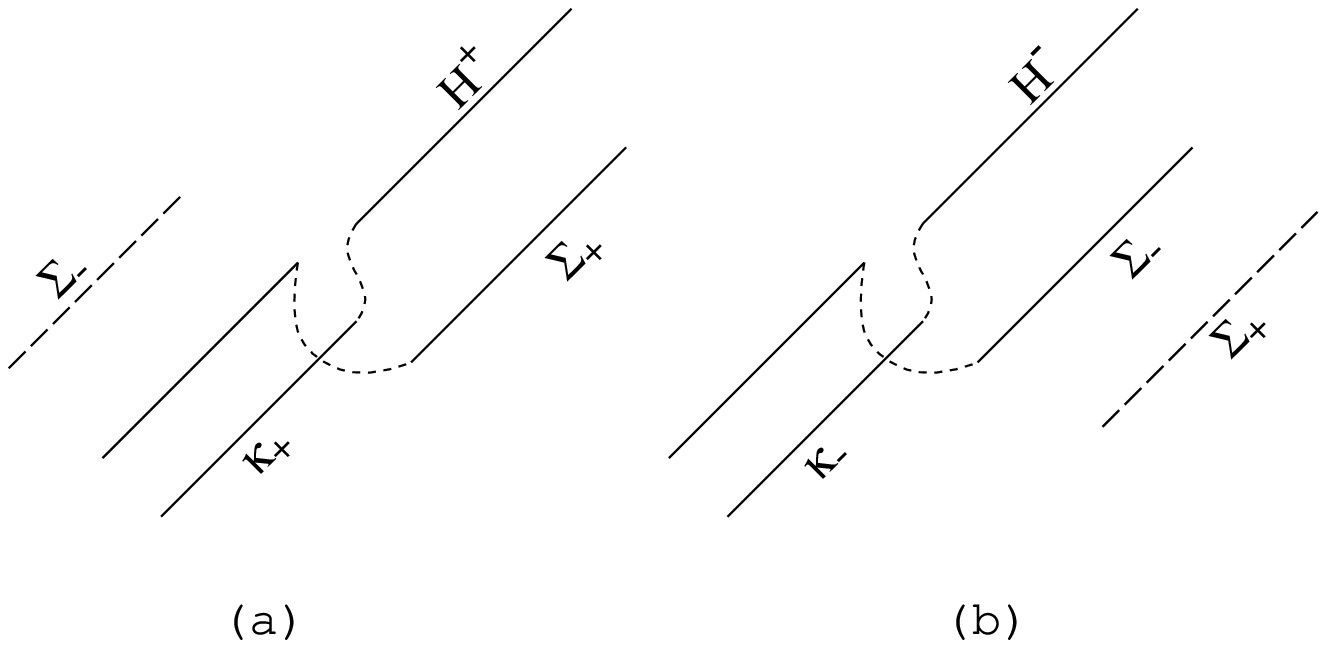}
\end{center}
\caption{Shell-tunneling through the future horizon. First (a), the outer face $\Sigma_+$ tunnels through the initial location $H^+$ of the
apparent horizon. Finally (b), the inner face $\Sigma_-$ tunnels through the shrunken horizon $H^-$.
}
\label{Fig3}
\end{figure}

\hspace*{6 mm}
In a real black hole formed by collapse, the L-sector of the L$\to$R tunneling scenario  
does not exist. A picture that corresponds more closely to our intuitive idea of Hawking evaporation
is shown in Figure \ref{Fig3}. 
A virtual particle, represented by a thin or thick spherical shell (outer 
and inner face histories $\Sigma_+$ and $\Sigma_-$), formed just
 beneath the future horizon $H^+$, tunnels through the horizon from sector F to sector R and escapes.
The attendant  loss of mass causes the horizon to shrink from $H^+$ to $H^-$.
(In Figs \ref{Fig3}(a) and (b) the two face histories $\Sigma_+$ and $\Sigma_-$ are shown separately for clarity.)

Using this overall picture, we shall here derive two formulae for the tunneling probability $\gamp(E)$, based on
the thin- and thick-shell models, and follow up with a discussion. Both formulae appear in the literature,
sometimes in one and the same paper. In the simplest case of Schwarzschild evaporation the two formulae are equivalent,
but they differ in general.

For optimum tunneling probability, the outward pre-tunneling velocity should be as large, and post-tunneling velocity as small (compatible with marginal escape from the horizon) as possible. This requires that both pre- and post-tunneling histories should be very nearly surfaces of constant retarded time $u$. Tunneling thus takes place between two given values of $u$, inside and outside that horizon, where
\[
u=-\frac{1}{\kappa_\epsilon}\ln U_\epsilon \quad (\epsilon=\pm)
\]
and $\kappa_+$ and $\kappa_-$ are the surface gravities before and after the shell has escaped. According to (\ref{elcct15f}), in the process $\textrm{Im }u$ jumps from $\frac{\pi}{2\kappa_+}$ before the horizon shrinks to  $-\frac{\pi}{2\kappa_-}$ afterwards.

(One way to visualize the tunneling for a thin shell is to imagine the pre- and post- tunneling histories reslotted so that the  $U_+=0$ and $U_-=0$ axes are aligned. This permits both histories to be mapped onto a single Kruskal diagram.)

In the Appendix it is noted that Schwarzschild time $t$ may be subjected to arbitrary space-dependent translations
\begin{equation}\label{elcct17s}
t \to t_{gen}=t+\psi(r)
\end{equation}
without affecting the validity of the expression for the Jacobi action
\begin{equation}
W=-\int Md\tau + \int \dbar E \int dt_{gen}(E)
\label{elcct18}
\end{equation}
or the definition of conjugate variable $E$, the energy measured at infinity. (We have slightly extended the thin-shell expression (\ref{A23}) so as to be applicable also to thick shells with layers of energy $\dbar E$ -- not an exact differential in general.)
where, as in (\ref{elcct15w}), the tunneling energy $\dbar E$ is given by the
(inexact for $e\neq 0$) differential form
\begin{equation}\label{elcct21}
\dbar E=dm-\frac{e}{r_H}de
\end{equation}

For the tunneling episode, the natural choice of parameter is $t_{gen}=u$. This yields for the imaginary part of the thin-shell action
\begin{equation}\label{elcct22}
\textrm{Im } W_{thin}=E_{thin}\textrm{Im } \Delta u=E_{thin}\overline{\Big(\frac{\pi}{\kappa}\Big)}
\end{equation}
and for a thick shell,
\begin{equation}\label{elcct20}
\textrm{Im }W_{thick}=\int \dbar E\frac{\pi}{\kappa(E)}
\end{equation}
where $\kappa(E)$ is the surface gravity after mass $E$ has been lost by the hole. (Along the pre- and post- tunneling segments, one could take
$t_{gen}=v$. They, of course, do not contribute to $\textrm{Im } W$.
)

We can now go on to consider in turn the thin- and thick- shell tunneling models.

For a \emph{thin shell}, we obtain from (\ref{elcct15w}), and the work of the Appendix,
\[
E_{thin}=-\left(\Delta m-\overline{\Big(\frac{e}{r_H}\Big)}\Delta e\right)
\]
where $\Delta m, \Delta e$ are the (negative) changes in the hole's mass and charge.

This yields the tunneling probability
\begin{equation}\label{elcct23}
\gamp_{thin}(E)=e^{-2\textrm{Im }W}=\exp \left(-\overline{\Big(\frac{2\pi}{\kappa}\Big)} E_{thin}\right),
\end{equation}
in agreement with our result (\ref{elcct17}) for L$\to$R thin-shell tunneling. We note that $E_{thin}$ can be written more simply:
\[
E_{thin}=-\Delta\Big(m-\frac{1}{2}\frac{e^2}{r_+}\Big)=-\frac{1}{2}\Delta r_+.
\]

In the case of a \emph{thick shell}, we are faced with the inexactness of $\dbar E$ when $e\neq 0$,
which makes the line integral (\ref{elcct20}) path-dependent and non-unique. If, however,
we assume further that the evaporation is, to a sufficient approximation, quasi-stationary
 -- i.e.,
that we can neglect dissipative effects like gravitational radiation as the hole settles
into its new configuration -- then it follows from the first law of black hole mechanics
\cite{gid11} that
\begin{equation}\label{elcct24}
-2\pi\dbar E/\kappa(E)=\frac{1}{4}dA=dS
\end{equation}
the (negative) changes of horizon area and Bekenstein-Hawking entropy, which \emph{are} exact
differentials. 

The integral (\ref{elcct20}) now gives exactly and unambiguously
\begin{equation}\label{elcct25}
\gamp(\Delta m, \Delta e)=e^{\Delta S}
\end{equation}

Because this derivation does not rely on spherical symmetry in any essential way,
(\ref{elcct25}) extends readily to the evaporation of holes with spin. One then replaces
(\ref{elcct24}) with the extended first law \cite{gid11}
\begin{equation}\label{elcct26}
\frac{1}{4}dA=\frac{2\pi}{\kappa}(dm-\Phi_H de-\omega_H dJ)
\end{equation}
where $J$ is the hole's angular momentum and $\omega_H$ the angular velocity of the horizon.
We forgo further details here, since the extension is straightforward.
(In contrast, the corresponding extension of the thin-shell formula(\ref{elcct22}) encounters
the difficulty that it is generally impossible to achieve isometry of the two faces of a thin
shell sandwiched between different Kerr geometries.)

In (\ref{elcct23}) and (\ref{elcct25}) we have two different formulae, based on a thin-shell and
a continuum model respectively, for the emission probability $\gamp$ from a charged black hole.
Explicitly, the exponents are
\begin{equation}\label{elcct27}
-\overline{\Big(\frac{2\pi}{\kappa}\Big)}E=\pi\overline{\Big(\frac{r^2_H}{\sqrt{m^2-e^2}}\Big)}\Delta r_H, \quad
\Delta S=2\pi\overline{r_H}\Delta r_H
\end{equation}
for the thin- and thick-shell cases respectively.
For uncharged black holes these two expressions agree, but they begin to diverge for non-vanishing
charge.

It will require a deeper investigation to decide which of these expressions is more
correct, or (more likely) to bring to light some more complex formula which amalgamates features 
of both. Both depend only, as they should, on the observable, i.e. the states before and after
emission.

One expects discontinuous aspects of the quantized emission to be more marked at low temperatures,
and here the thin-shell formula (\ref{elcct23}), which predicts zero emission probability for zero 
temperature ($\kappa_+=0$), accords well with expectations. In general, however, it is not clear
how well a thin-shell idealization is able to handle the complex form of the Einstein-Maxwell 
interaction term $ede/r_H(m,e)$ in (\ref{elcct21}). 
The arithmetic-mean prescription (\ref{A18})
for the thin-shell action was designed to deliver the right classical equations of motion 
(\ref{A19}) for the
shell, but that depends on the simple bilinear form $q\varphi$ of the interaction (\ref{A16}). 
Arithmetic-mean recipes are generally limited to linear theories (like Einstein's in the
distributional limit -- only terms linear in second metric-derivatives survive in the
curvature) and bilinear interactions. 
On these grounds we tend to favour the continuum-model
formula (\ref{elcct25}) over (\ref{elcct23}) for charged black holes whose surface gravity is 
not too small.
It is questionable 
whether naturally evaporating black holes ever approach a state of extremality 
(stable or unstable); 
numerical evidence suggests that angular momentum and charge are evaporated preferentially. 
All of this raises questions worth exploring further, in particular the speculative possibility 
of stable, information-storing black hole remnants.

\section{Correlations, entropy, unitarity}\label{sbk5}
\hspace*{6 mm}
We found in the previous section that the probability of a black hole of mass $m$ and charge
$e$ (hence Bekenstein-Hawking entropy $S(m,e)=\frac{1}{4}A_H$) decaying to a state $(m_1,e_1)$
is
\begin{equation}\label{elcct61}
\gamp (m,e\to m_1,e_1)=e^{-(S-S_1)}
\end{equation}
according to a thick-shell model for the  evaporating particle. (On the alternative thin-shell model,
(\ref{elcct61}) becomes an inequality, $P< e^{-(S-S_1)}$.) The result is obviously transitive and leads to
\[
\gamp(m,e\to m_1,e_1\to m_2,e_2\to\cdots\to 0)=e^{-S(m,e)}
\]
for any chain of decays ending in complete evaporation. If $N(m,e)$ is the number of ways the hole can
evaporate, we must have $N\gamp=1$. Hence
\begin{equation}\label{elcct62}
N=e^S
\end{equation}
This gives an interpretation of the Bekenstein-Hawking entropy in terms of the number of modes of 
evaporation. Equation (\ref{elcct61}) can now be reinterpreted as
equating, to leading order (up to a neglected cross-section factor), the
transition probability to a statistical factor $e^{S_1}/e^S$, equal to the ratio of
the number of final states to initial states.

Thus, the degrees of freedom in the outgoing radiation equal (or even exceed, on the thin-shell model)
the maximum information capacity of the hole, as measured by the Bekenstein-Hawking entropy. This 
provides evidence, purely on the basis of counting, that unitarity could be preserved and that the 
radiation has enough room to accommodate all of the information. (But this has not been a matter of 
universal agreement\cite{gid3}.)

In a significant paper \cite{gid4}, Zhang et al have reopened the question whether deviations from 
exact thermality encoded in (\ref{elcct23}) or (\ref{elcct61}) produce correlations actually capable
of carrying information.

Generally, one defines the correlation coefficient $C(a,b)$ between two events $a$ and $b$ by
\begin{equation}\label{elcct63}
C(a,b)=\ln\frac{P(a,b)}{P(a)P(b)}
\end{equation}
where $P(a,b)$ is the probability of both $a$ and $b$, and $P(b)=\sum_a P(a,b)$ the probability of 
$b$. The \emph{conditional} probability of $b$ (given that $a$ has already occurred) is
\begin{equation}\label{elcct64}
P(b|a)=\frac{P(a,b)}{P(a)}
\end{equation}

In our case (confining the argument to uncharged evaporation for simplicity), the probability that
a black hole of mass $m$ emits a quantum of energy $E$ is, by (\ref{elcct61}),
\[
\gamp(m,E)=\exp\Big\{ -4\pi[m^2-(m-E)^2] \Big\}
\]
and this yields a nontrivial correlation \cite{gid4}
\begin{equation}\label{elcct65}
C(E_1,E_2)=\ln\frac{P(m,E_1+E_2)}{P(m,E_1)P(m,E_2)}=8\pi E_1 E_2
\end{equation}

It might be thought that one should replace $P(m,E_2)$ in (\ref{elcct65}) by $P(m-E_1,E_2)$ to 
take account of mass loss from the first emission. That would be tantamount to replacing $P(b)$
in (\ref{elcct63}) by the conditional probability $P(b|a)$. But one sees at once from (\ref{elcct64})
that this would give $C(a,b)=0$ identically for \emph{any} two events. The argument is clearly circular:
it absorbs the correlations themselves into the test for their existence.

Thus, as emphasized in \cite{gid4}, the original conclusion (\ref{elcct65}) is correct. The 
radiation does have correlations and, according to (\ref{elcct62}), these have the capacity to carry
off the maximum information content of the hole.

\section{Summary and conclusions.}\label{sbk6}
\hspace*{6 mm}
The argument for information loss in black hole evaporation rested, first and foremost, on the belief 
that the emerging spectrum is exactly thermal. Since 1994 this main pillar of the argument has toppled.
A series of investigations has shown that, when back-reaction of the emission is allowed for, there
are departures from thermality\cite{gid2}, and that the associated correlations have the capacity to
carry off the maximum information content of the hole \cite{gid4}. In this paper we have reviewed, 
simplified and somewhat extended this work.

A logically prior issue, not touched on in this work, is the extent to which the emission process
can access and extricate information buried in the depths of the hole. The wavelength and period of
the radiation when it reaches infinity are of the order of the hole's diameter and characteristic 
internal time
(i.e. maximum proper free-fall time from the horizon), suggesting that access is possible. On the
other hand, the original mode analyses,
(non-local, but still the best evidence for thermality across the spectrum) 
and the usual particle tunneling and pair-creation arguments suggest a
localized, near-horizon process with high frequencies at formation that are subsequently redshifted.
This would make the emission an exclusively surface process, without access to the interior.
However, the uncertainty principle forbids such definite allocation of the place and time
of emission for a quantum of given energy. It cannot be ruled out that much of the radiation
reaches infinity without substantial redshift, having been formed by concerted, coherent action
of the  entire hole. The issue clearly needs further study (cf \cite{gid12}) but may well turn out
to be a red herring as a possible culprit for information loss.
\\

Steve Carlip kindly informs us that equation (\ref{elcct25})
was previously obtained by Massar and Parentani \cite{gid13}, using
path-integral methods developed by Carlip and Teitelboim.

We thank Samir Mathur, Douglas Singleton and the referee for correspondence and helpful suggestions.

\appendix
\section{Review of shell dynamics}\label{sdya}
\hspace*{6 mm}
This appendix collects and reviews the origins of the shell formulae used in section \ref{sbk4}.
More detail can be found in chapter 3 of Poisson's \emph{Toolkit} \cite{gid7}.

The shell history is a timelike 3-space $\Sigma$ 
that divides spacetime into two sectors
$\mathcal{V_+}$ and $\mathcal{V_-}$ mapped by independent charts $x^\alpha_+$ and $x^\alpha_-$ and with metrics 
\[
ds^2_{\pm}=g_{\alpha\beta}dx^\alpha dx^\beta\lvert_{\pm}
\]
(Greek indices run from 1 to 4).  Their common boundary $\Sigma$ is described by two
sets of imbedding relations $x^\alpha_\pm=x^\alpha_\pm(\xi^a)$ and intrinsic 3-metric
\[
ds^2=g_{ab}(\xi)d\xi^a d\xi^b
\]
in terms of a third independent set of intrinsic coordinates $\xi^a$
(Latin indices run from 1 to 3). The intrinsic 3-metric $g_{ab}$ is induced 
compatibly on $\Sigma$ by each of the two 4-geometries via
\begin{equation}
g_{ab}(\xi)=\Big(g_{\alpha\beta}(x)e^\alpha_{(a)}e^\beta_{(b)}\Big)_\pm,\qquad
e^\alpha_{(a)}(\xi)\lvert_\pm=\frac{\partial x^\alpha_\pm}{\partial\xi^a}
\end{equation}
where $e_{(a)}$ are three basis vectors tangent to $\Sigma$.

The interior and exterior imbeddings $(ds)^2_{\Sigma_+}=(ds)^2_{\Sigma_-}$ thus
induce a unique tangential 3-metric $g_{ab}$ on $\Sigma$;
However, its extrinsic curvatures
\begin{equation}
K^\pm_{ab}=n_{\alpha|\beta}e^\alpha_{(a)}e^\beta_{(b)}\lvert^\pm
\end{equation}
in $\mathcal{V_+}$ and $\mathcal{V_-}$ will generally differ in the presence of a shell of
matter.(Here, $n_\alpha$ is the unit spacelike normal, directed from $\mathcal{V_-}$ to
$\mathcal{V_+}$ and the stroke denotes 4-dimensional covariant differentiation;
3-dimensional covariant derivatives with respect to $g_{ab}$ are indicated by a semi-colon)

The jump of extrinsic curvature across $\Sigma$, 
\[
[K_{ab}]\equiv K^+_{ab}-K^-_{ab}
\]
gives the shell's intrinsic stress-energy tensor $S_{ab}$ via a surface analogue
of Einstein's field equations:
\begin{subequations}\label{A1}
\begin{equation}\label{A1a}
-8\pi S_{ab}=[K_{ab}-g_{ab}K]
\end{equation}
\begin{equation}\label{A1b}
S^b_{\phantom{2}a;b}=-[e^\alpha_{(a)}T^\beta_\alpha n_\beta]
\end{equation}
\end{subequations}
Equation (\ref{A1b}) is an energy conservation law, describing the shell's response
to the stresses and energy fluxes in its surroundings.
\\

Variation of the metric in the action
\[
I=I_{geom}+I_{mat}
\]
where
\begin{equation}\label{A2}
16\pi I_{geom}=\int_{bulk}d^4x\sqrt{-g}R^\alpha_\alpha -2\int_\Sigma [K]d\Sigma
\end{equation}
yields, beside Einstein's field equations, also the jump conditions (\ref{A1a})
Here, the shell is treated simply as an isometric
 pair of timelike boundaries
$\Sigma_+$ and $\Sigma_-$ of the bulk, with a common normal $n$, directed from
$\mathcal{V_-}$ to $\mathcal{V_+}$.

The key geometrical result needed in the derivation is that the variation of
\begin{equation}\label{A3}
I=\int_{bulk}d^4x\sqrt{-g}R^\alpha_\alpha +\int_{boundary}2\epsilon K d\Sigma
\end{equation}
is
\begin{equation}\label{A4}
\delta I=\int_{bulk} G^{\alpha\beta}\delta g_{\alpha\beta}\sqrt{-g}d^4x
+\int_{boundary}\pi^{ab}\delta g_{ab}d\Sigma
\end{equation}
where $\pi_{ab}=-\epsilon(K_{ab}-g_{ab}K)$, $\epsilon=n\cdot n=\pm 1$ and
$n$ is now the outward normal to the bulk, bounded by a timelike or spacelike $\Sigma$.
An apparent sign discordance between (\ref{A2}) and (\ref{A3}) merely reflects these
differing conventions for the orientation of $n$.

Variation of the material part of the action $I_{mat}$ gives the stress-energy
tensor of the shell and its environment according to
\begin{equation}\label{A5}
\delta I_{mat}=\int\frac{1}{2}T^{\alpha\beta}\delta g_{\alpha\beta}\sqrt{-g}d^4x
+\int\frac{1}{2}S^{ab}\delta g_{ab}d\Sigma
\end{equation}
Vanishing of (\ref{A4})+(\ref{A5}) gives the Einstein field equations $G^{\alpha\beta}=8\pi T^{\alpha\beta}$ and the shell equation (\ref{A1a}).

For a shell of fluid in a fluid environment, $I_{mat}$ takes the form
\begin{equation}\label{A6}
I_{mat}=-\int_{bulk}\rho(n,s)\sqrt{-g}d^4x-\int_\Sigma \sigma(n,s)d\Sigma
\end{equation}
where $\rho$ and $\sigma$ are energy densities per unit volume and unit area respectively,
 and $s, n$ the corresponding entropy and molecular number densities.
(Our use of the same symbols $n, s$ for densities per unit volume and unit area in these different contexts should
not cause confusion.)

Our specific concern in section \ref{sbk4} is with a class of spherical shells moving in spherical geometries of the form
\begin{equation}\label{A10}
(ds^2)_{\pm}=\frac{dr^2}{f(r)}+r^2d\Omega^2-f(r)dt^2
\end{equation}
with different functions $f_-$ and $f_+$ and time coordinates $t_-$ and $t_+$ in 
$\mathcal{V_-}$ and $\mathcal{V_+}$. In all geometries of the form (\ref{A10}), the Einstein field equations
require $T^r_r=T^t_t$, which implies that $e^\alpha_{(a)}T^\beta_\alpha n_\beta=0$ and $S^b_{a;b}=0$ at 
the shell by virtue of (\ref{A1b}). 
Thus in this class of spacetimes, the ambient pressures do no work
and the shell's internal energy is conserved.

The junction and conservation laws(\ref{A1}) now reduce to a single equation of motion for the shell radius $R(\tau)$:
\begin{subequations}\label{A11}
\begin{equation}\label{A11a}
[\mE]=-\frac{M}{R}
\end{equation}
\begin{equation}\label{A11b}
dM+Pd(4\pi R^2)=0
\end{equation}
\end{subequations}
where $M=4\pi R^2\sigma$ is the shell's proper mass, $\sigma=-S^\tau_\tau$ and $P=S^\theta_\theta=S^\phi_\phi$
is the surface pressure. The expression (\ref{A11a}) follows at once from
\[
K^\theta_\theta=\frac{1}{2}n^\alpha\partial_\alpha\ln g_{\theta\theta}=n^r/R= \mE/R \qquad (\pm\text{understood})
\]
where 
\begin{equation}\label{A12}
\mE=n\cdot\nabla r=ft_\tau=\eta\sqrt{f+R^2_\tau}
\end{equation}
The sign factor $\eta=\pm 1$ determines whether $r$ is increasing or decreasing outwards
(i.e, in the direction from $\mathcal{V_-}$ to $\mathcal{V_+}$), and the subscript $\tau$ indicates $\frac{d}{d\tau}$.
(Physically, $\mathcal{E_-}$ and $\mathcal{E_+}$ could be interpreted as energies of test particles of unit rest mass
attached to the inner and outer shell faces.)

The rationalized version of (\ref{A11a}) is 
\begin{equation}\label{A13}
R^2_\tau=\Big(\frac{m_\Sigma}{M}\Big)^2-\overline{f}(R)+\frac{1}{4}\Big(\frac{M}{R}\Big)^2
\end{equation}
The bar denotes an arithmetical mean $\overline{f}=\frac{1}{2}(f_++f_-)$ and we have defined the Schwarzschild mass
$m(r)$ interior to radius $r$ by
\[
f(r)=1-\frac{2m(r)}{r}, \quad m_\Sigma =m_+(R)-m_-(R)
\]
The simplest illustration is a shell of mass $m$ in empty space:
(\ref{A11}) or (\ref{A13}) yields 
\[
m=M\sqrt{1+R^2_\tau}-\frac{M^2}{2R}
\]
which is a decomposition of the total (conserved) mass-energy $m$ in a form whose Newtonian counterpart is 
self-evident.
\\

These results extend readily to charged shells.
We simply augment the action (\ref{A2})+(\ref{A6}) with an interaction term
\begin{equation}\label{A14}
I_{int}=\int J^\alpha A_\alpha \sqrt{-g}d^4x
\end{equation}
plus the usual free-field Lagrangian $-\frac{1}{16\pi}F_{\alpha\beta}F^{\alpha\beta}$.
Our spherical constraint admits only radial electric fields, for which $A_\alpha$ is gauge-reducible in 
static $(r,t)$ coordinates to a single component
\begin{equation}\label{A15}
A_t=-\varphi=-\frac{e_\pm}{r}
\end{equation}
in $\mathcal{V_\pm}$ respectively, up to an additive constant which could be adjusted across the shell. The shell's charge is $q=e_+-e_-$ and (\ref{A14}) reduces to
\begin{equation}\label{A16}
I_{int}=\int qA_\alpha dx^\alpha
\end{equation}
Adding (\ref{A16}) to the action does not affect the junction conditions or equations of motion in the forms
(\ref{A1}), (\ref{A11}) or (\ref{A13}).(They were obtained by varying the metric in the action; however, (\ref{A16}) is
independent of the metric.) Electric forces are nonetheless now at work in these equations, for instance through
the Faraday stresses now present in $T^\beta_\alpha$ on the right-hand side of (\ref{A1b}).
\\

Our assumption of spherical symmetry has reduced the effective dynamics to one degree of freedom -- the shell radius $R$.
We recall that the shell was introduced in the first place in Sec. \ref{sbk4} as an s-wave approximation to the wave function 
for \emph{particles} tunneling radially, so that a (1+1)-dimensional description is entirely apposite.
However, the action that we have is still (3+1) dimensional, and saddled with elements, such as a
3-dimensional extrinsic curvature, foreign to a (1+1)-dimensional dynamics. It is therefore appropriate
to seek a (1+1)-dimensional effective Lagrangian which delivers the same equations of motion for R and
whose Hamiltonian generates the corresponding quantum evolution.

A Lagrangian of this sort emerges if we retain just the mechanical, non-geometrical parts
(\ref{A6})+(\ref{A14}) of the original action:
\begin{equation}
I_{eff}=-\int Md\tau +\int q A_\alpha dx^\alpha
\label{A17}
\end{equation}
To fix the Lagrangian we still need to choose a time-coordinate invariant under Euler-Lagrange variation
of the particle's world-line.(Proper time $\tau$(=arc length) is clearly inadmissible) As we shall see,
a formally successful(as well as physically desirable) choice is a static observer's time $t$, split
evenly between the alternatives $t_+$ and $t_-$ by taking the arithmetic mean of the + and - actions.
This prescription leads to the following explicit form of (\ref{A17}):
\begin{equation}
I_{eff}=\int\overline{Ldt}, \qquad L=-M\frac{f}{\mE}-q\varphi
\label{A18}
\end{equation}
The upper bar denotes an arithmetic mean, $\overline{X}\equiv\frac{1}{2}(X_++X_-)$; $\varphi_{\pm}=e_{\pm}/R-c_\pm$,
where $c_\pm$ are adjustable constants,
and we have used $d\tau=(fdt/\mE)_{\pm}$, which follows from (\ref{A12}).

The Euler-Lagrange equation $\delta I_{eff}/\delta R=0$ for (\ref{A18}) gives
\begin{equation}\label{A19}
R_{\tau\tau}\overline{\Big(\frac{1}{\mE}\Big)}+\frac{1}{2}\overline{\Big(\frac{f'(R)}{\mE}\Big)}
+\frac{\overline{\mE}}{M}\frac{dM}{dR}-\frac{q}{MR^2}\overline{e}=0
\end{equation}
Where prime indicates that it is a differentiation with respect to $R$.

On the other hand, the junction condition (\ref{A11a}) can be written in either
of the forms ($\epsilon=\pm$)
\begin{equation}\label{A19a}
\frac{2M}{R}\mE_\epsilon =f_--f_+-\epsilon\frac{M^2}{R^2}
\end{equation}
Addition yields
\begin{equation}\label{A19b}
\frac{2M}{R}\overline{\mE}=f_--f_+
\end{equation}
Differentiation of this equation yeilds a second order equation of motion identical with (\ref{A19}), when $f$ has the Reissner-Nordstr\"{o}m form.

Thus (\ref{A18}) is indeed the desired \emph{effective particle action} for the shell.
\\

The canonical radial momentum and Hamiltonian associated with Lagrangian (\ref{A18}) are
\begin{eqnarray}
p_r&=&\frac{\partial L}{\partial R_t}=\frac{M}{f}R_\tau\nonumber\\
H&=&p_rR_t-L=M\mE+q\varphi\label{A20}
\end{eqnarray}

$\overline{H}$ is conserved by Hamilton's equations for a time-independent background. 
Explicit evaluation for Reissner-Nordstr\"{o}m fields, using definition (\ref{A12}), shows that
\begin{equation}\label{A21}
\overline{H}\equiv E=m-\overline{c}
\end{equation}
where $m=m_+-m_-$, $\overline{c}=q\overline{e}/r_H$
equal to the shell's Schwarzschild mass.
It is, incidentally, straightforward to verify that  (\ref{A17}) is the action, and (\ref{A20}) the Hamiltonian,
for a charged test \emph{particle} moving radially in the spherical geometry (\ref{A10}).

The effective action (\ref{A18}) is taken between given endpoints$(R_i,t_i) (i=1,2)$. Under a change of endpoints, the change
of extremal action is given by the Hamilton-Jacobi formula
\begin{equation}
\delta I_{eff}=\left[\overline{p}\delta R-\overline{H\delta t} \right]^2_1=[p\delta R]-E(\delta t_2-\delta t_1)
\label{A22}
\end{equation}
(where we have assumed $\delta t_+=\delta t_-$)

Our interest in section \ref{sbk3} et seq is in the probability for a particle of given \emph{energy} $E$ to tunnel
between given points $R_1$ and $R_2$ straddling a horizon;  the tunneling time is irrelevant. The path
which maximizes this probability is the path which minimizes the Jacobi action.
\begin{equation}
W\equiv I_{eff}+\int Edt
\label{A23}
\end{equation}
for which (\ref{A22}) gives
\begin{equation}
\delta W=[p\delta R+ t\delta E]=[p\delta R]+ (t_2-t_1)\delta E
\end{equation}

As is evident from this form of the Hamilton-Jacobi formula, the Hamiltonian, and hence also the Jacobi action (\ref{A23})
(with the same energy $E$) are invariant under space-dependent translations of the time coordinate. Thus, (\ref{A23}) remains 
valid if $t$ is replaced, for example,  
by retarded or advanced times $u,v$ (See (\ref{elcct14r})), or by any of a class of generalized times $t_{gen}$, defined by
\[
dt_{gen}=dt\pm\alpha(f)\frac{dr}{f},
\]
in which $\alpha$ has an arbitrary dependence on $f(r)$ subject to $\alpha(0)=1$.
That includes  Painlev\`{e}-Gullstrand time, defined by
\[
dt_{PG}=dt+\sqrt{1-f}\frac{dr}{f},
\]

\end{document}